\documentclass[
preprint,
 amsmath,amssymb,
 aps,
prb,
]{revtex4-1}

\usepackage{graphicx}
\usepackage{dcolumn}
\usepackage{bm}
\usepackage[version=3]{mhchem} 
\usepackage[T1]{fontenc}       
\usepackage{graphicx}
\usepackage[amssymb]{SIunits}
\usepackage{amssymb}
\usepackage{ulem}
\usepackage{textcomp}
\usepackage{braket}
\usepackage{upgreek}
\usepackage{xcolor}
\usepackage{comment}

\begin{document}


\title{Controlling a nuclear spin in a nanodiamond}

\author{Helena S. Knowles}
\affiliation{Cavendish Laboratory, University of Cambridge, JJ Thomson Ave, Cambridge CB3 0HE, UK}

\author{Dhiren M. Kara}
\affiliation{Cavendish Laboratory, University of Cambridge, JJ Thomson Ave, Cambridge CB3 0HE, UK}

\author{Mete Atat\"{u}re}
\affiliation{Cavendish Laboratory, University of Cambridge, JJ Thomson Ave, Cambridge CB3 0HE, UK}
\email{ma424@cam.ac.uk}
\date{\today}

\begin{abstract}
The sensing capability of a single optically bright electronic spin in diamond can be enhanced by making use of proximal dark nuclei as ancillary spins. Such systems, so far only realized in bulk diamond, provide orders of magnitude higher sensitivity and spectral resolution in the case of magnetic sensing, as well as improved readout fidelity and state storage time in quantum information schemes. In nanodiamonds, which offer additional opportunities as mobile nanoscale sensors, electronic-nuclear spin complexes have remained inaccessible. We demonstrate coherent control of a $^{13}$C nuclear spin located 4\,\AA\ from a nitrogen-vacancy center in a nanodiamond and show quantum-state transfer between the two components of this hybrid spin system. We extract a nuclear-spin free precession time of  $T_2^* = 26\,\upmu$s, which exceeds the bare electron free precession time in nanodiamond by two orders of magnitude.
\end{abstract}

\pacs{Valid PACS appear here}
\maketitle

{\it Introduction ---} \rm{The} nitrogen-vacancy center (NV) in diamond is one of the most versatile optically active spins in a solid-state system. Its electronic spin has successfully been used as a long-lived qubit for quantum information schemes\cite{deLange2010,Bar-Gill2013} and as a magnetic field detector with single-spin sensitivity \cite{Grinolds2013,Grinolds2014,Muller2014}. There have been a number of reports demonstrating the benefits of including ancillary nuclear spins to this device for quantum state storage and error correction, particularly important for implementing quantum computing architectures\cite{Childress2006,Dutt2007,Neumann2010,Robledo2011,Maurer2012,Waldherr2014,Pfaff2014,Reiserer2016}. For sensing schemes the advances in exploiting ancilla spins have been more recent, showing orders of magnitude improvement in sensitivity and spectral resolution using the host N nuclear spin\cite{Laraoui2013a,Lovchinsky2016,Zaiser2016,Rosskopf2016,Haberle2016,Pfender2016}. These enhancements result from the complementary properties of electronic and nuclear spins: The electronic spin provides optical access and high sensitivity to weak magnetic fields, while the nuclear spin provides a good quantum memory due to its weak coupling to the environment. This allows a coupled electron-nuclear system to outperform its single-spin constituents.

NVs in nanodiamonds enable nanoscale measurements inside liquids, soft tissue and even living cells\cite{McGuinness2011,Kucsko2013,Neumann2013}. They have also recently been used in combination with a scanning probe to image the motion of magnetic domain boundaries\cite{Tetienne2014}. However, they exhibit poor coherence times compared with NVs in bulk diamond ($T_2 \sim \upmu$s in nanodiamonds versus ms in bulk diamond at room temperature). This is a consequence of typically high paramagnetic impurity concentrations, as well as proximity of the NV to surfaces. Hybrid electronic-nuclear spin schemes are, therefore, of particular interest for nanodiamonds, as they could provide access to long nuclear coherence times and lifetimes, leading to up to three orders of magnitude improvement in spectral resolution of magnetic field sensing\cite{Laraoui2013a}. However, utilizing the host N as ancilla spin requires a strong magnetic field and high-power radio frequency pulses, which limits the capacity of the NV sensor.

In this Article, we work with a $^{13}$C isotope nuclear spin, a spin-$1/2$ particle, and investigate its potential as an ancilla for NV-based magnetic field sensing in a nanodiamond. We demonstrate a tuneable coupling between the NV and the $^{13}$C spin, which allows fast quantum gates between the two species without the requirement of direct nuclear spin driving or strong static magnetic fields. The control of the nuclear-spin state is achieved through nuclear-spin-selective rotations of the NV spin combined with free evolution times matched to the nuclear spin precession period. Further, we polarize the proximal nuclear spin and probe its free precession time, T$_2^*$, finding it to be two orders of magnitude longer than that of the NV spin.

\vspace{0.7cm}
\begin{figure}[h]
\centering
\includegraphics[width=0.8\textwidth]{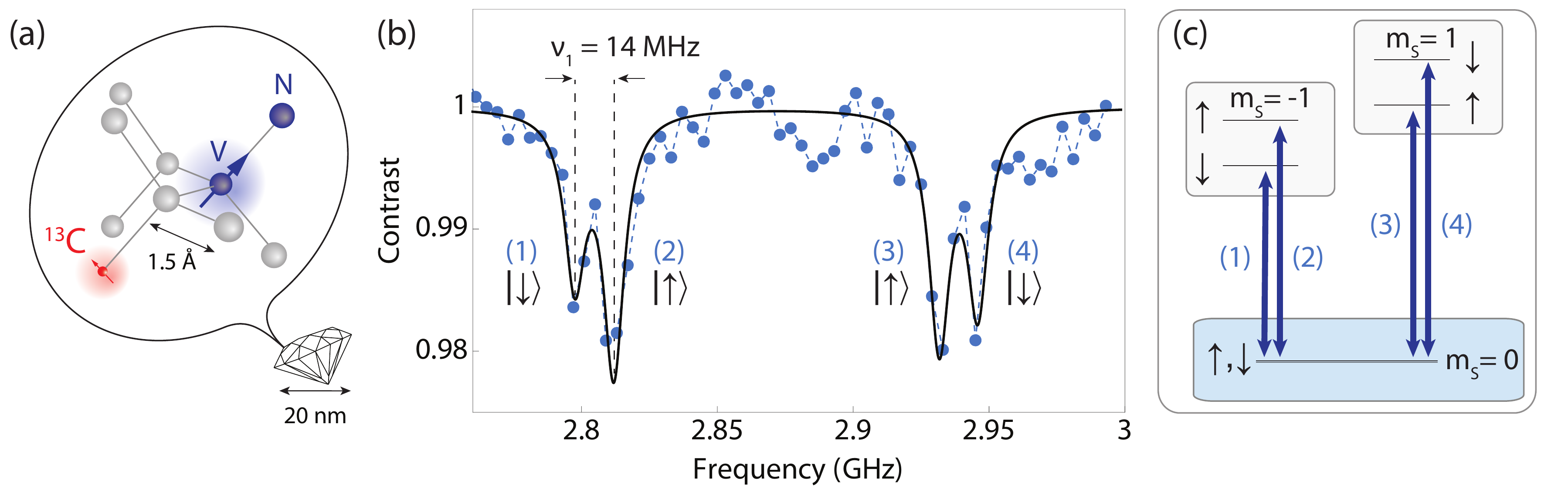}
\caption{(a) Illustration of a possible arrangement of the coupled spin system inside the nanodiamond. (b) Continuous wave ODMR spectrum of a NV spin coupled to a single proximal $^{13}$C nuclear spin. The solid line is a fit to a sum of four Lorentzian resonances and we extract a hyperfine-induced splitting $\nu_1 = 14.3 \pm 1$\,MHz. (c) Energy level diagram showing the levels of the coupled system with NV electron spin states $m_S = 0,\pm1$ and $^{13}$C nuclear spin states $\ket{\uparrow}$ and $\ket{\downarrow}$. Nuclear-spin conserving electronic spin transitions are labeled (1)-(4) and each correspond to a resonance in panel (c).}
\label{13CODMR}
\end{figure}

{\it Coupled nuclear-electron spin pair ---} {\rm The} nanodiamonds used in this work have a natural $^{13}$C isotope concentration of 1.1\%. This results in an approximately 1-nm mean separation between the $^{13}$C atoms in the diamond lattice. These atoms are randomly placed with respect to the NV and on average 1 in 3 NVs possesses at least one strongly coupled $^{13}$C atom with $A^{^{13}\mathrm{C}}>1/T^{*}_{2,\rm{NV}}$, where $A^{^{13}\mathrm{C}}$ is the hyperfine coupling and $T^{*}_{2,\rm{NV}}$ is the NV coherence time\cite{Smeltzer2011}. In order to identify NVs with proximal $^{13}$C atoms, as illustrated in Fig.\ 1(a), we look for the signature hyperfine splitting of the optically detected magnetic resonance (ODMR) signal of the NV spin. The NV-$^{13}$C hyperfine interaction leads to two resonances for each of the $\ket{m_S=0} \rightleftharpoons\ket{m_S=\pm1}$ NV transitions corresponding to the $\ket{m_I=\uparrow}$ and $\ket{m_I=\downarrow}$ states of the nuclear spin. Figure 1(b) displays the resulting ODMR spectrum with 14-MHz splitting, where the nuclear-spin-conserving transitions are labeled (1) to (4). This interaction leads to a hybrid energy level scheme as shown in Fig.\ 1(c).

The Hamiltonian for the NV ground state in the presence of one proximal $^{13}$C atom in units of frequency is given by
\begin{equation}
\hat{\mathrm{H}} = D_{\rm{gs}} \hat{\rm{S}}_z^2 + \frac{\upmu_{\rm{e}}}{h} {\boldsymbol{B}} \cdot{\hat{\mathrm{\mathbf{S}}}} + \frac{\upmu_{\rm{n}}}{h}  {\boldsymbol{B}}\cdot {\hat{\mathrm{\mathbf{I}}}}+\sum\limits_{\mu,\nu = \rm{x,y,z}} \alpha_{\rm{\mu \nu}} \hat{\mathrm{S}}_{\rm{\mu}} \hat{\mathrm{I}}_\nu,
\label{Ham13C}
\end{equation}
where $\upmu_{\mathrm{e}} = g_{\rm{e}} \cdot \upmu_{\mathrm{B}} = 1.9 \cdot 10^{-23}$ J/T and $\upmu_{\mathrm{n}} = 3.5 \cdot 10^{-27}$ J/T are the NV electronic and $^{13}$C nuclear magnetic moments, respectively, $D_{\mathrm{gs}} = 2.87$\,GHz is the NV ground state splitting, $\boldsymbol{B}$ is the external magnetic field ($|\boldsymbol{B}| \sim$ 7\,mT in our case) and $ \hat{\bf{S}} = (\mathrm{\hat{S}}_x,\mathrm{\hat{S}}_y,\mathrm{\hat{S}}_z)$ and  $\hat{\bf{I}} = (\mathrm{\hat{I}}_x,\mathrm{\hat{I}}_y,\mathrm{\hat{I}}_z)$ are the spin operator vectors given by the Pauli operators as $ \hat{\bf{S}} =  \hat{\bf{I}} = \frac{1}{2}(\hat{\sigma}_x,\hat{\sigma}_y,\hat{\sigma}_z)$.
The component of the hyperfine interaction that results in the observed ODMR splitting shown in Fig. 1(b) is given by 
\begin{equation}
A^{^{13}\rm{C}} = \sum\limits_{\nu = \rm{x,y,z}}\alpha_{\rm{z \nu}}.
\end{equation}
Different possible locations of the $^{13}$C atoms in the diamond lattice lead to discrete values for this coupling, reaching over 130 MHz when the $^{13}$C lies on a neighbouring site to the NV\cite{Jelezko2004,Felton2009}. The coupling strength for this NV is 14.3$\,\pm 1\,$MHz, which corresponds to one of 9 lattice sites at a separation of about 4\,\AA\ between the $^{13}$C and the NV spin\cite{Gali2009,Smeltzer2011}. 

\it{Nuclear spin dynamics ---} \rm{The} $^{13}$C spin experiences a local magnetic field that is dependent on the state of the NV spin: The electron states $\ket{\pm1}$ result in a large dipole field ${\boldsymbol{B}}_{\pm\rm{1}}^{\rm{dip}}$ at the site of the nuclear spin and the nuclear Zeeman interaction is negligible in comparison. Consequently, the ${\boldsymbol{B}}_{\pm\rm{1}}^{\rm{dip}}$ magnitude and direction determine the energy splitting, which we term $h\nu_{\pm1}$, (the ODMR splitting of each electron resonance in Fig. 1(b)) and the quantization axis of the $^{13}$C, respectively. When ${\boldsymbol{B}}$ is aligned along the NV axis and the NV is in the $\ket{0}$ state there is no dipole field, i.e. ${\boldsymbol{B}}_{0}^{\rm{dip}}=0$, and the external field dictates  the energy splitting $h\nu_{0}=\upmu_{\rm{n}}\, B_{\rm{z}}$ (not resolvable in Fig. 1(b)) and quantization axis of the $^{13}$C. 
If ${\boldsymbol{B}}$ is not aligned with the NV axis, the electronic spin eigenstates are mixed as $\ket{0}\rightarrow\ket{0'}=\alpha\ket{0}+\beta\ket{1}+\gamma\ket{-1}$, where $\alpha\gg\beta,\gamma$. Due to the finite $\ket{\pm1}$-character of $\ket{0'}$ the $^{13}$C nuclear spin experiences a non-zero dipole field ${\boldsymbol{B}}_{\rm{0'}}^{\rm{dip}}$. This modifies the $^{13}$C precession frequency from $\nu_{0}$ to $\nu_{0'}$. 
We calculate this modified frequency by treating the non-secular terms (terms containing $\hat{\mathrm{S}}_x$ and $\hat{\mathrm{S}}_y$) of the Hamiltonian in Eq.~1 as a perturbation to the secular terms. The perturbation is given by
\begin{equation}
\hat{\rm{H}}_\perp = \frac{\upmu_{\rm{e}}}{h}(B_{\rm{x}}\hat{\mathrm{S}}_{\rm{x}}+B_{\rm{y}}\hat{\mathrm{S}}_{\rm{y}}) + \sum\limits_{\nu = \rm{x,y,z}}(\hat{\mathrm{S}}_{\rm{x}}\alpha_{\rm{x \nu}}+\hat{\mathrm{S}}_{\rm{y}}\alpha_{\rm{y\nu}})\hat{\mathrm{I}}_\nu,
\end{equation}
where $(B_x,B_y,B_z)$ are the components of the external magnetic field, ${\boldsymbol{B}}$. 
The new eigenenergies for the electronic and nuclear states $m_S$ and $m_I$, respectively, can be calculated from 
\begin{equation}
\begin{split}
E_{(m_S,m_I)}^{(2)} = E_{(m_S,m_I)}^{(0)} + \sum \limits_{k \neq m_S,l \neq m_I} \frac{\langle k, l | H_\perp |k, l \rangle} {E_{m_S,m_I}^{(0)}-E_{k,l}^{(0)}}.
\end{split}
\end{equation}
Since $\alpha_{\upmu\nu}^2 \ll (\upmu_eB_\upmu/h)^2, \alpha_{\upmu\nu} \upmu_eB_{\upmu}/h$ for all $\upmu,\nu$ at the magnetic fields we apply, we neglect any terms of order $\alpha^2$. The enhanced precession frequency of the $^{13}$C spin is then given by
\begin{equation}
\nu_{0'} = \frac{\upmu_n |\boldsymbol{B}|}{h} +\frac{4 \sqrt{2} \upmu_e}{h} \frac{B_x \alpha_{xz} + B_y \alpha_{yz}}{D_{gs}},
\end{equation}
where we have taken the limit $D_{\mathrm{gs}} \gg \upmu_\mathrm{e} B_z /h, \alpha_{zz}$. The correction term scales as $\alpha_\perp (\upmu_e B_\perp)/hD_{\mathrm{gs}}$ and the nuclear precession frequency is thus enhanced by
\begin{equation}
\propto \frac{\alpha_\perp}{D_{\mathrm{gs}}} \frac{\upmu_e B_\perp}{\upmu_n |\boldsymbol{B}|},
\end{equation}
with $B_\perp = \sqrt{B_x^2+B_y^2}$, $\alpha_\perp = \sqrt{\alpha_{xz}^2+\alpha_{yz}^2}$. For hyperfine interactions of only a few MHz this factor can make $\nu_{0'}$ significantly larger than $\nu_{0}$, since $\upmu_{\rm{e}} \sim 5000\,\upmu_{\rm{n}}$. This allows for a $\nu_{0'}$ that is tunable through the orientation and the magnitude of the external magnetic field: the stronger the non secular component $B_\perp$ compared to the secular component $B_z$, the higher $\nu_{0'}$. The correction to the nuclear spin energies, $h\nu_{\pm1}$ when the NV is in the $\ket{\pm1'}$ states, scales as $\alpha_\perp (\upmu_e B_\perp/h)^2/D_{gs}^2$. This term is negligible compared to other energy scales of the system and therefore we use the unperturbed value, $h \nu_{\pm1}$.

We can observe the enhanced precession of the $^{13}$C spin by using the NV as a probe of the oscillating magnetic field created by the precessing nuclear spin. The data in Fig. 2(a) show a spin echo decay with an external magnetic field orientated 45 degrees from the NV axis. 
We scan $\tau$ and observe oscillations in the echo signal. These correspond to a nuclear spin precession of $\nu_{0'} = 1.6$\,MHz, which is more than two orders of magnitude higher than the bare Larmor frequency, $\upmu_n \lvert \boldsymbol{B} \rvert/h = 5.9$~kHz at $\lvert \boldsymbol{B} \rvert = \rm{7\,mT}$. Spin echo data with a field aligned along the NV axis are shown in Fig. 2(b). Here, the contrast decays monotonously with a decay constant of 1.5\,$\upmu$s and displays no oscillations, confirming the tuneability of $\nu_{0'}$. 

\begin{figure}[h]
\centering
\includegraphics[width=0.6\textwidth]{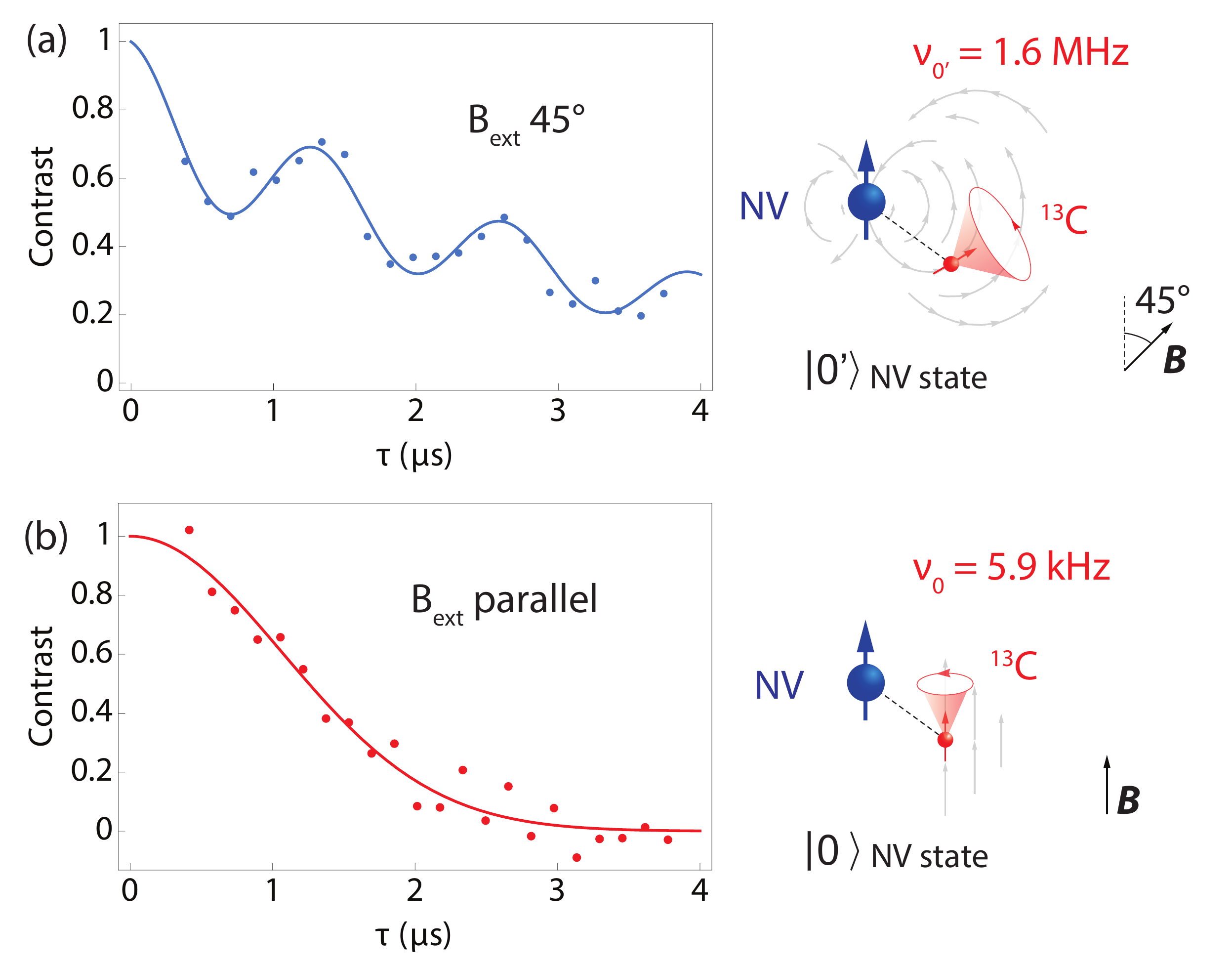}\\
\caption{NV spin echo decay in a magnetic field that is (a) at 45\degree\ and (b) parallel to the NV axis. A fit to $\mathrm{exp}(-\tau/T_{\mathrm{SE}})\mathrm{cos}(2\pi \nu_0 \tau/2)$ yields the enhanced $^{13}$C precession frequency of $\nu_{0'} = $ 1.6\,MHz in (a).}
\label{13Cspinecho}
\end{figure}

\it{Proximal nuclear spin polarisation ---} \rm{Through} the application of sequential pulsed optical spin preparation and microwave (MW) rotation of the NV spin, we can make use of the enhanced precession of the $^{13}$C spin to initialize it. We define $\ket{\uparrow}$ and $\ket{\downarrow}$ as the eigenstates of the nuclear spin when the NV is in the $\ket{1}$ state. In the $\ket{0'}$ NV state however, the $^{13}$C spin has a different quantization axis and the nuclear eigenstates are given by $\ket{+} = a \ket{\downarrow} + b e^{i \phi}\ket{\uparrow}$ and $\ket{-} = b \ket{\downarrow} - a e^{i \phi} \ket{\uparrow}$, where $a$ and $b$ are real numbers and $a^2 + b^2 = 1$. If $a = b$, these quantization axes are perpendicular, enabling full polarization transfer between the NV and the $^{13}$C. To understand the $^{13}$C polarization scheme, we consider an initially unpolarized $^{13}$C spin and the special case of $a = b$. In the $\{\ket{\uparrow},\ket{\downarrow}\}$ basis, we first consider the $\ket{\uparrow}$ component. After optical initialization of the NV we have the state $\ket{0',\uparrow}$. A MW $\pi$-pulse with Rabi frequency $\Omega <\nu_1$ but $\Omega >\nu_{0'}$ is applied resonant with the $\ket{0',\pm}\leftrightarrow\ket{1,\downarrow}$ transition and thus does not drive the state $\ket{0',\uparrow}$. Because $\ket{0',\uparrow}$ is not an eigenstate, it precesses freely into $\ket{0',\downarrow}$ and back at a frequency$\nu_{0'}$. This precession is illustrated in the upper Bloch sphere in Fig. 3(a). After a free precession time $\tau_{^{13}\mathrm{C}} = 1/2\nu_{0'}$, we achieve a successful polarization into the target state, $\ket{0',\downarrow}$, verified by ODMR readout. We next consider the effect of the protocol on the $\ket{\downarrow}$ component. After NV initialization the $\pi$-pulse rotates the electron spin yielding the state $\ket{1,\downarrow}$. The nuclear spin remains stationary in this eigenstate. NV re-initialization after $\tau_{^{13}\mathrm{C}}$ creates again the state $\ket{0',\downarrow}$. Thus, after a free evolution time of $\tau_{^{13}\mathrm{C}}$ the $^{13}$C spin is driven into the $\ket{\downarrow}$ state by the polarization scheme, regardless of its initial state. For the general case the maximal polarization that can be reached is given by $p_\mathrm{max} = 1 - (a^2 - b^2)^2$, where $p_\mathrm{max} = 0,1$ for no polarization and full polarization, respectively. This case is illustrated in the lower Bloch sphere in Fig. 3(a), where the angle $\delta$ between the two quantization axes can take any value between 0 and 90\degree.

\begin{figure}[h]
\centering
\includegraphics[width=0.6\textwidth]{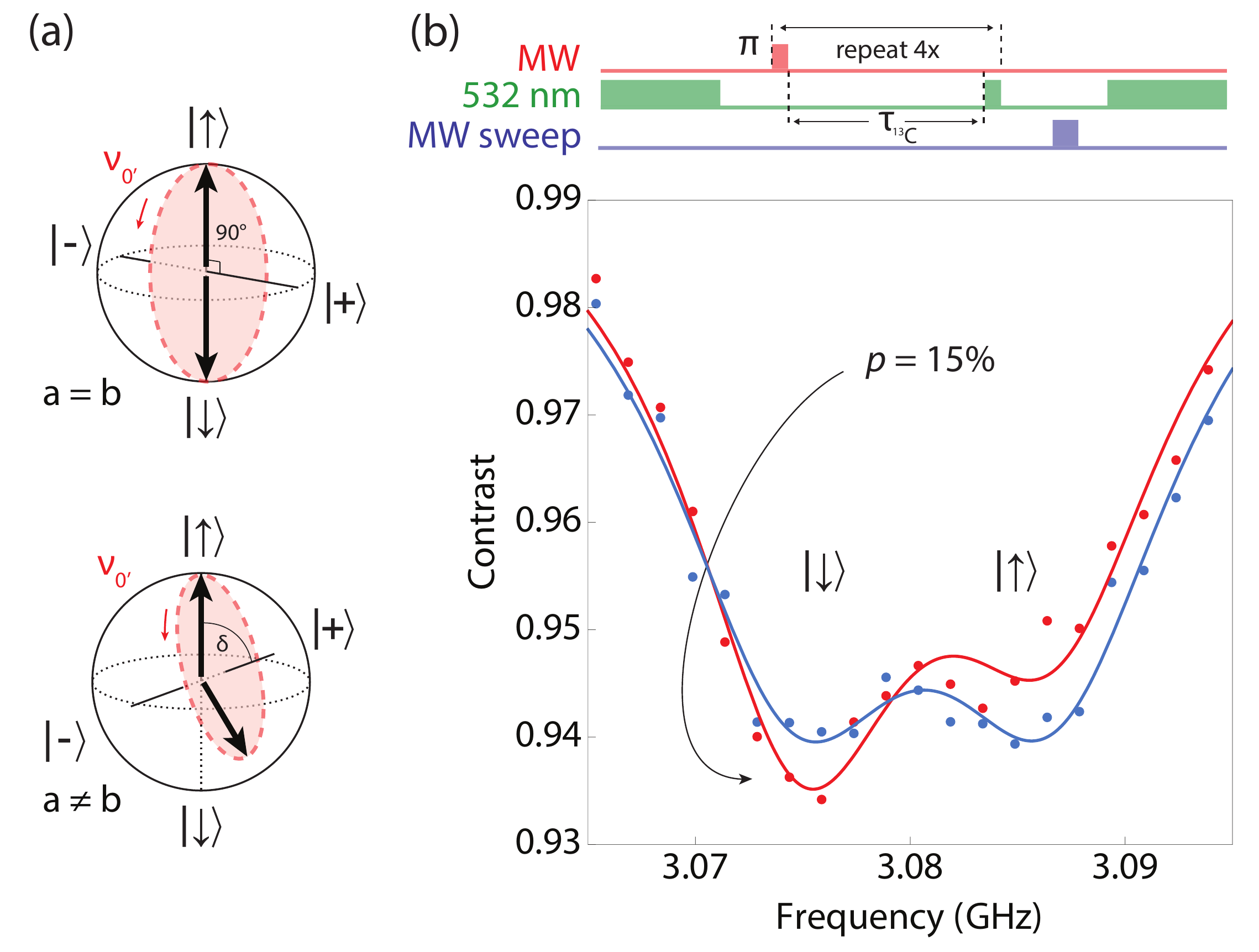}
\caption{(a) Bloch spheres illustrating the eigenbases of the nuclear spin when the NV is in the $\ket{1}$ state ($\ket{\uparrow}$ and $\ket{\downarrow}$) and when the NV is in the $\ket{0'}$ state ($\ket{+}$ and $\ket{-}$). The angle between the two quantization axes is 90\degree\ and $0< \delta <$ 90\degree\ in the top and bottom sphere, respectively. (b) Pulsed ODMR measurement with (blue circles) and without (red circles) the nuclear spin polarization sequence. Both measurements are fit to a double Lorentzian resonance and we extract a polarization as defined in the text of $p = 15$\%.}
\label{13Cpolarization}
\end{figure}

If we start the sequence with an initialized electron spin and a pure nuclear spin state, $\ket{0'} \otimes (p\ket{\downarrow}+q\ket{\uparrow})$, this scheme can also be viewed as quantum state transfer between the $^{13}$C state and the NV state: After the selective $\pi$ pulse and the $\tau_{^{13}\mathrm{C}}$ wait, this state evolves into $(p\ket{1}+q\ket{0'}) \otimes \ket{\downarrow}$ and the original nuclear spin state is transferred onto the electron. Starting with an initialized nuclear spin and a superposition state for the NV electron spin allows the inverse mapping process. This offers the ability to store and retrieve quantum states in the $^{13}$C spin.

Figure 3(b) shows the sequence used to polarize and then measure the state of the nuclear spin. Four polarization steps are performed, followed by a final MW pulse, which is scanned in frequency (blue pulse trace) and provides probing of the nuclear spin state. The blue circles represent ODMR data in the absence of the polarization scheme and show two resonances with equal amplitude for each of the nuclear spin states. The red circles show polarization into the $\ket{\downarrow}$ state. We fit the data to a double Lorentzian dip and extract the amplitudes of both resonances $p_{\downarrow}$ and $p_{\uparrow}$. These reveal the population of the nuclear states $\ket\downarrow$ and $\ket{\uparrow}$, respectively. The degree of nuclear spin polarization is then given by
\begin{equation}
p = \frac{p_{\downarrow}-p_{\uparrow}}{p_{\downarrow}+p_{\uparrow}}
\end{equation}
and we extract a polarization of $p = 15$\%, corresponding to a spin temperature of 110\,mK, achieved at room temperature. 

Ideally, $p_\mathrm{max}$ should be reached in one polarization step. In practice, however, several repetitions of the sequence are needed to create an observable polarization. This is due to the long ($>$100 ns), $\pi$-pulse durations required to resolve the $\ket{0'\pm}\leftrightarrow\ket{1,\downarrow}$ and $\ket{0'\pm}\leftrightarrow\ket{1,\uparrow}$ transitions. During these pulses the nuclear spin rotation at $\nu_{0'}$ is not negligible and hence the rotation is not an ideal $\pi$-pulse. The efficiency of this polarization scheme could therefore be improved by adjusting $\boldsymbol{B}$ to reduce $\nu_{0'}$ and using shaped NV $\pi$-pulses to create a narrow output frequency distribution whilst maintaining high Rabi rates. The fundamental limit, however, is the relative orientation of the two NV-state dependent $^{13}$C quantization axes. Assuming perfect rotations of the NV spin, our degree of polarization corresponds to an angle of 67\degree\ between the two quantization axes.

\it{Storage time of a quantum state by a nuclear spin ---} \rm{To} quantify the quantum state storage time of a $^{13}$C spin in a nanodiamond, we apply the pulse sequence shown at the top of Fig. 4: The NV spin is initialized into the $\ket{0'}$ state followed by a $\pi$ pulse on the $\ket{0'\pm}\leftrightarrow\ket{1,\downarrow}$ resonance. After a free interaction time $\tau$ another $\pi$ pulse at the same frequency is performed, inverting the population of the NV spin only when the nuclear spin is in the $\ket{\downarrow}$ state. Following the second $\pi$ pulse the NV state is measured. If the nuclear spin is initially in the $\ket{\uparrow}$ state it precesses freely in the enhanced external field during $\tau$. This results in a probability of measuring the bright NV state $\ket{0'}$ of $P_{\ket{0'}} = |a^2+b^2 e^{\mathrm{i}2\pi\nu_{0'}\tau}|^2$. If it was in the $\ket{\downarrow}$ state, the NV is rotated by the first $\pi$ pulse into the $\ket{1}$ state, where the nuclear state $\ket{\downarrow}$ is an eigenstate and does not evolve in time. This leads to a bright state NV population independent of the waiting time $\tau$ and results in $P_{\ket{0'}} = 1$ after the second $\pi$-pulse. Hence, the sequence results in an average signal of $P_{\ket{0'}} = (1+a^4+b^4 + 2a^2b^2\mathrm{cos}(2\pi\nu_{0'}\tau))/2$. This is presented in Fig. 4, where we show data taken at $\boldsymbol{B}$ aligned almost parallel to the NV axis. We see oscillations at $\nu_{0'} = 170$ kHz, arising from the free precession of the nuclear spin. These oscillations decay with a characteristic time of $T_2^* = 26\,\upmu$s, the nuclear spin coherence time. This represents a 100-fold enhancement in the storage time of a quantum state by the nuclear spin in comparison to using only the electron spin with its coherence time of a few hundred ns\cite{Knowles2014}. This storage time could be further enhanced by using a spin echo sequence on the nuclear spin\cite{Dutt2007}.

\begin{figure}[h]
\centering
\includegraphics[width=0.5\textwidth]{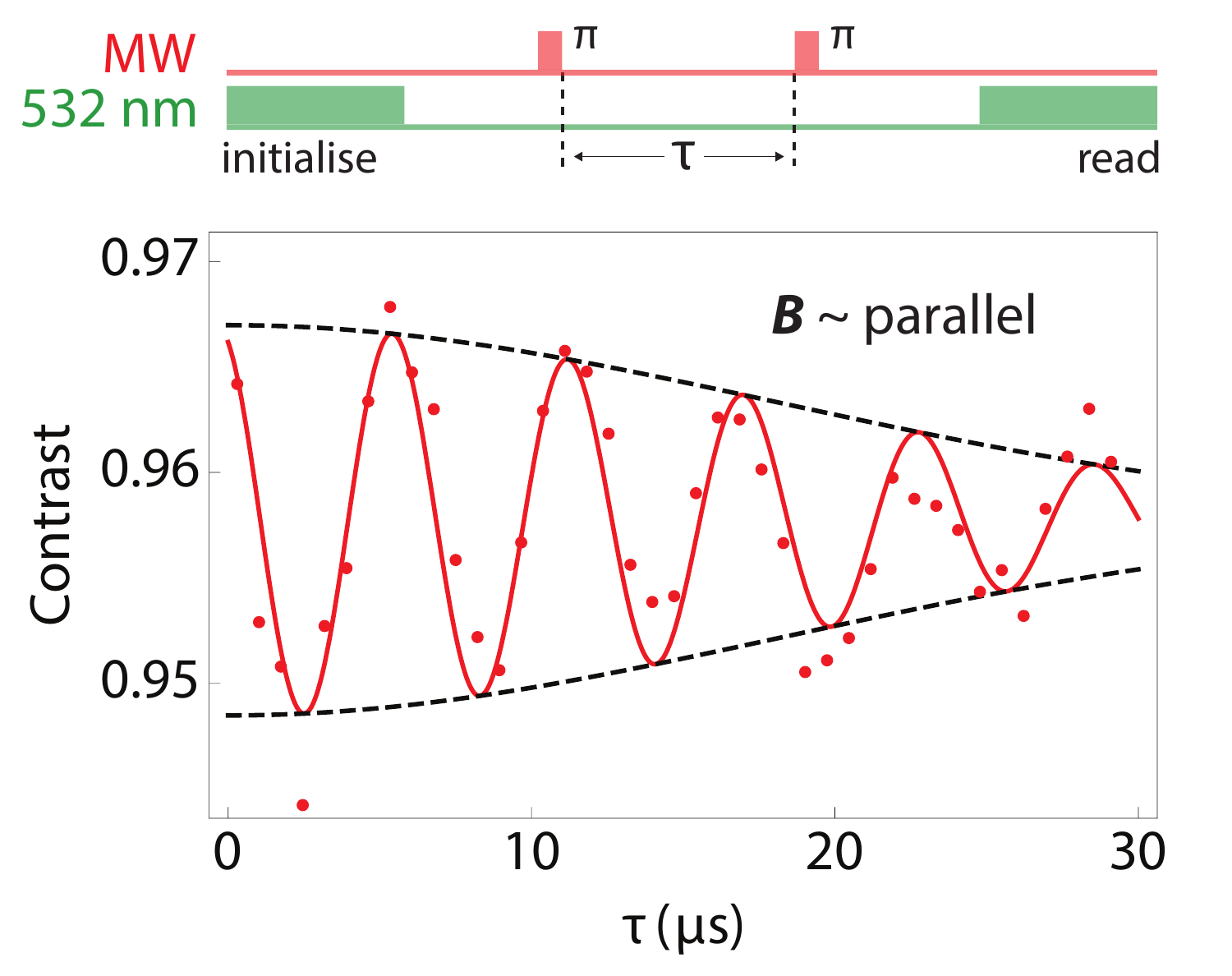}
\caption{Probing the coherence time of the nuclear spin. The solid curve is a fit proportional to $\mathrm{cos}(2\pi\nu_{0'}\tau) \mathrm{exp}[-(\tau/T_2^*)^2]$ and the dashed curves show the envelope of the fit. Here, $\nu_{0'} = 170 $\,kHz and the coherence time, $T_2^*$, is 26 $\upmu$s.}
\label{13Cpi}
\end{figure}

We demonstrate coherent manipulation of a nuclear spin and its tunable coupling to the NV electronic spin. This enables the initialization of the nuclear spin and access to long-time storage of arbitrary quantum states in a nanodiamond. So far, ancilla-enhanced sensors have relied on the host N nuclear spin of the NV. Our results with a proximal $^{13}$C offer advantages for non-invasive sensing, as we require neither a $\gtrsim$0.5\,T static magnetic field, to suppress NV-N spin exchange, nor high-power radio frequency (RF) excitation for coherent nuclear spin rotations. Consequently, this system is well suited for the investigation of novel materials, where NV scanning probes are a unique tool offering the ability to probe nanoscale magnetic order and phase transitions. Also, the absence of high-power RF has the benefit of avoiding excessive heating when working with biological samples. For both sensing applications, nuclear spins open up the exciting prospect of measuring spin and charge dynamics on the 0.1 to 10 ms timescales, inaccessible to the NV electronic spin alone. Considering that nanodiamonds represent the extreme of proximity to surfaces, the results in this work show that ancilla-based protocols used in bulk diamond may also be realized in nanoscale structures, such as optical cavities and waveguides\cite{Momenzadeh2015,Li2015,Sipahigil2016}, in the context of on-chip quantum photonics devices.

\begin{acknowledgments}
We gratefully acknowledge financial support by the Leverhulme Trust Research Project Grant 2013-337, the European Research Council ERC Consolidator Grant Agreement No. 617985 and the Winton Programme for the Physics of Sustainability. H.S.K. also acknowledges financial support by St John's College through a Research Fellowship. We thank Jean-Fran\c{c}ois Roch for stimulating discussions.
\end{acknowledgments}


%

\end{document}